\documentclass[aps,pre,reprint,superscriptaddress]{revtex4-1}

\pdfoutput=1

\usepackage{graphicx}
\usepackage{amssymb}
\usepackage{amsmath}
\usepackage{xcolor}
\usepackage{pdfpages}
\usepackage{pgffor}
\usepackage{CJK}
\usepackage{multirow}
\usepackage{gensymb}

\makeatletter
\AtBeginDocument{\let\LS@rot\@undefined}
\makeatother

\begin{document}

\title{Modeling Solution Drying by Moving a Liquid-Vapor Interface: Method and Applications} 

\author{Yanfei Tang}
\affiliation{Department of Physics, Center for Soft Matter and Biological Physics, and Macromolecules Innovation Institute, Virginia Tech, Blacksburg, Virginia 24061, USA}
\author{John E. McLaughlan}
\affiliation{Department of Physics, Center for Soft Matter and Biological Physics, and Macromolecules Innovation Institute, Virginia Tech, Blacksburg, Virginia 24061, USA}
\author{Gary S. Grest}
\affiliation{Sandia National Laboratories, Albuquerque, NM 87185, USA}
\author{Shengfeng Cheng}
\email{chengsf@vt.edu}
\affiliation{Department of Physics, Center for Soft Matter and Biological Physics, and Macromolecules Innovation Institute, Virginia Tech, Blacksburg, Virginia 24061, USA}
\affiliation{Department of Mechanical Engineering, Virginia Tech, Blacksburg, Virginia 24061, USA}

\begin{abstract}
A method of simulating the drying process of a soft matter solution with an implicit solvent model by moving the liquid-vapor interface is applied to various solution films and droplets. For a solution of a polymer and nanoparticles, we observe ``polymer-on-top'' stratification, similar to that found previously with an explicit solvent model. Furthermore, ``polymer-on-top'' is found even when the nanoparticle size is smaller than the radius of gyration of the polymer chains. For a suspension droplet of a bidisperse mixture of nanoparticles, we show that core-shell clusters of nanoparticles can be obtained via the ``small-on-outside'' stratification mechanism at fast evaporation rates. ``Large-on-outside'' stratification and uniform particle distribution are also observed when the evaporation rate is reduced. Polymeric particles with various morphologies, including Janus spheres, core-shell particles, and patchy particles, are produced from drying droplets of polymer solutions by combining fast evaporation with a controlled interaction between the polymers and the liquid-vapor interface. Our results validate the applicability of the moving interface method to a wide range of drying systems. The limitations of the method are pointed out and cautions are provided to potential practitioners on cases where the method might fail.
\end{abstract}

\maketitle

\section{Introduction}

Drying is a phenomenon we witness everyday~\cite{Dincer2016}. It is also a useful tool for material fabrication~\cite{Brinker2004, Zhou2017AdvMater}. In a typical process, solutes are dissolved or dispersed in an appropriate solvent and the resulting solution is dried to yield desired materials or structures. For~example, drying is frequently used to produce paint coatings~\cite{KeddieRouth2010, Keddie1997, Luo2008, Kooij2015}, polymer films~\cite{Strawhecker2001, Luo2005, Sousa2019}, polymeric particles~\cite{Shin2017ACSNano, Zhu2021}, and polymer nanocomposites~\cite{Jouault2014, Imel2015, Cheng2016, Kumar2017JCP}. The drying characteristics of respiratory droplets plays an important role in determining the fate and transmissibility of respiratory viruses including the COVID-19 virus responsible for the ongoing pandemic~\cite{Ge2021, Leung2021}. Because of its practical importance and rich nonequilibrium physics~\cite{KeddieRouth2010, Routh2013}, many efforts have been devoted to elucidate the fundamental processes and mechanisms of drying for soft matter solutions~\cite{Zhou2017AdvMater, Schulz2018}, including molecular dynamics simulations~\cite{Tsige2004MMa, Tsige2004MMb, Tsige2005, Cheng2013, Cheng2016, Cheng2017, Fortini2016, Fortini2017, Howard2017, Howard2017b, Howard2018, Reinhart2017, Reinhart2018, Statt2017, Statt2018, Tatsumi2018, Tang2018Langmuir, Tang2019Langmuir_control, Tang2019JCP_compare}.

To model the drying process of a soft matter solution, a~key challenge is the treatment of the solvent. To~capture factors that may be important in an evaporation process, such as hydrodynamic interactions between solutes~\cite{Howard2018, Statt2018}, evaporation-induced flow in the solution (e.g., capillary flow in an evaporating droplet showing the coffee-ring effect)~\cite{Deegan1997}, and instabilities during drying including Rayleigh-B\'{e}nard and B\'{e}nard-Marangoni instabilities~\cite{Mitov1998, Strawhecker2001, Bassou2009, Chapman2021}, it is ideal to include the solvent explicitly in a computational model~\cite{Cheng2013, Cheng2016, Cheng2017, Tang2018Langmuir, Tang2019Langmuir_control, Tang2019JCP_compare, Reinhart2017, Reinhart2018, Howard2018, Statt2018}. However, such models are computationally extremely expensive as the solvent particles significantly outnumber the solutes at realistic volume fractions. Usually, systems containing several million particles or more have to be considered even for a few hundred nanoparticles~\cite{Cheng2012, Cheng2013}. Because of the large system size, only fast evaporation rates can be modeled using this approach~\cite{Cheng2013, Cheng2016, Cheng2017, Tang2018Langmuir, Tang2019Langmuir_control, Tang2019JCP_compare, Reinhart2017, Reinhart2018, Howard2018, Statt2018}. Considering the limitations of explicit solvent models, it is natural to explore alternative approaches that are computationally more efficient and able to quickly yield results that are at least qualitatively reasonable. One such effort is to model the solvent as an uniform and viscous medium in which the dispersed solutes are moving around. This approach leads to various implicit solvent models of soft matter solutions~\cite{Grest2011, Howard2018, Statt2018, Tang2019JCP_compare}.

Recently, implicit solvent models have been applied to study the evaporation process of particle suspensions and polymer solutions and reveal many interesting phenomena induced by drying~\cite{Fortini2016, Fortini2017, Howard2017, Howard2017b, Howard2018, Statt2017, Statt2018, Tatsumi2018, Tang2019JCP_compare}. Fortini~et~al. used Langevin dynamics simulations based on an implicit solvent model to study the drying of a bidisperse colloidal suspension film and demonstrated the counterintuitive ``small-on-top'' stratification in which the smaller particles are predominately distributed on top of the larger particles after drying~\cite{Fortini2016, Fortini2017}. Tatsumi~et~al. used a similar model to investigate the role of evaporation rates on stratification~\cite{Tatsumi2018}. Howard~et~al. and Statt~et~al. employed this approach to simulate the drying of colloidal suspensions~\cite{Howard2018}, colloidal mixtures~\cite{Howard2017}, polymer-colloid mixtures~\cite{Howard2017b}, polymer-polymer mixtures~\cite{Howard2017b,Statt2018}, and polydisperse polymer mixtures~\cite{Statt2017}, and observed stratifying phenomena as well. Recently, we demonstrated that comparable stratification could be observed for colloidal suspensions in both explicit and implicit solvent models that are carefully matched~\cite{Tang2019JCP_compare}.

In the implicit solvent approach to modeling drying, all the solutes are confined by a potential barrier, which represents the confinement effect of the liquid-vapor interface between the solvent and its vapor. One simple form of the confining potential is a harmonic potential. In~a previous work, Tang and Cheng analyzed the capillary force experienced by a small spherical particle at a liquid-vapor interface and showed that this harmonic approximation is reasonable for many situations~\cite{Tang2018PRE}. A rigorous physical foundation was thus established for the implicit solvent approach. In~this paper, we review the general method of using an implicit solvent model to study the drying process of a soft matter solution. A~careful implementation of the model has removed certain undesirable effects occurring at the liquid-vapor interface in several previous studies~\cite{Fortini2016, Howard2017, Howard2017b}. We further apply the method to various systems including solution films and droplets. Our results indicate that this approach can be applied to solutions with a good solvent where the solutes are initially well dispersed. Through the current study, the~advantages and possible deficiencies of the implicit solvent approach are revealed and~summarized.

The paper is organized as follows. In~Section~\ref{sec:ch7_ms}, the model and simulation method are introduced. The~applications of the method to various systems are presented in Section~\ref{sec:ch7_applications}. We briefly summarize the method and findings in Section~\ref{sec:ch7_summary}.

\section{Model and Simulation~Methodology} \label{sec:ch7_ms}

In an implicit solvent model, the~solvent is treated as a uniform, viscous, and~isothermal background~\cite{Tang2019JCP_compare}. The motion of a particle in this background is typically described by a Langevin equation that includes Stokes' drag as a damping term. The~damping rate can be chosen according to the effective viscosity of the solvent, the~diffusion coefficient of the particle, and~the particle size. To~model the liquid-vapor interface, a~potential barrier is used to confine all the particles in the liquid phase. As~shown previously~\cite{Tang2018PRE}, a harmonic potential can be employed as the barrier. The~evaporation process of the solvent, during~which the liquid-vapor interface recedes, can be mimicked by moving the location of the confining potential's minimum. Below~we discuss these two main ingredients of the implicit solvent approach of modeling solution~drying.

\subsection{Langevin~Dynamics} \label{ss:ch7_langevin}

In an implicit solvent, the~equation of motion of particles is given by the following Langevin Equation \cite{Schneider1978, Grest1986},
\begin{equation} \label{eq:ch7_langevin}
m \frac{d^2 \mathbf{r}_i}{dt^2} = \sum_{j \neq i} \mathbf{f}_{ij} + \mathbf{F}_i^D + \mathbf{F}_i^R
\end{equation}
where $m$ is the mass of the $i$-th particle, $\mathbf{r}_i$ is its position vector, $t$ is time, $\mathbf{f}_{ij}$ is the force on the particle from its interaction with the $j$-th particle, $\mathbf{F}_i^D$ is Stokes' drag, and~$\mathbf{F}_i^R$ is a random force. Stokes' drag can be expressed as
\begin{equation} \label{eq:ch7_fd}
\mathbf{F}_i^D = - \xi \frac{d \mathbf{r}_i}{d t}
\end{equation}
where $\xi$ is the friction (drag) coefficient of the particle. To~maintain the system at a constant temperature $T$, the~random force needs to follow the constraint set by the fluctuation-dissipation theorem,
\begin{eqnarray} \label{eq:ch7_fr}
	\begin{array}{c}
\langle \mathbf{F}_i^R(t) \rangle = 0 \\
\langle \mathbf{F}_i^R(t) \cdot \mathbf{F}_j^R(t') \rangle = 6 k_B T \xi \delta_{ij} \delta(t-t')
	\end{array}
\end{eqnarray}
where $\langle \cdot \rangle$ stands for an ensemble average, $k_\text{B}$ is the Boltzmann constant, $\delta_{ij}$ is the Kronecker delta, and~$\delta(t-t')$ is the Dirac delta~function.

In the dilute limit, the~diffusion coefficient $D$ of the particle is related to the friction coefficient $\xi$ through
\begin{equation} \label{eq:ch7_einstein}
D = \frac{k_B T}{\xi}
\end{equation}
which is known as the Einstein relation. For~a small particle with radius $R$ in a flow with a low Reynolds number, Stokes' law states that
\begin{equation} \label{eq:ch7_stokes_law}
\xi = 6 \pi \eta R
\end{equation}
where $\eta$ is the viscosity of the fluid. This yields the Stokes--Einstein relation, 
\begin{equation} \label{eq:ch7_stokes_einstein}
D = \frac{k_B T}{6 \pi \eta R}
\end{equation}
The friction coefficient $\xi$ is usually written in terms of a damping time $\Gamma$ as $\xi \equiv m/\Gamma$. As~a result, Stokes' drag becomes
\begin{equation}  \label{eq:ch7_fd_lammps}
\mathbf{F}_i^D = - \frac{m}{\Gamma} \frac{d \mathbf{r}_i}{d t}
\end{equation}
If Stokes' law holds, then $\Gamma = m/(6 \pi \eta R)$. If~we further assume the particle has a uniform mass density $\rho$, then $m= \frac{4}{3}\pi R^3 \rho$ and the damping time $\Gamma$ becomes
\begin{equation}  \label{eq:ch7_damping}
\Gamma = \frac{2\rho}{9 \eta} R^2
\end{equation}
The implication of this relationship is that for a bidisperse mixture of particles of size ratio $\alpha \equiv R_l/R_s$. where $R_l$ is the radius of the larger particles and $R_s$ is the radius of the smaller particles, the~damping time of the larger particles should be $\alpha^2$ times of that of the smaller ones in order for the Stokes--Einstein relation to hold for~both.

\subsection{Moving Interface~Method} \label{ss:ch7_moving_interface}

When the solvent evaporates from a solution, the~liquid-vapor interface recedes. To~mimic this process, the~location of the minimum of the potential barrier that is used to confine all the particles (in general, solutes) in the liquid solvent is moved toward the solution phase at speed $v_e$. For~evaporation at a constant rate, the~position of the liquid-vapor interface at time $t$ is given by $H(t) = H(0) - v_e t$, where $H(0)$ is the initial thickness of a suspension film or the initial radius of a spherical droplet. Similarly, $H(t)$ is the thickness of a drying film or the radius of a drying droplet at time $t$. For~a particle whose center is within distance $R$ from the liquid-vapor interface, where $R$ is a radius parameter, the~particle experiences a force normal to the interface. Otherwise, the~particle does not interact with the interface. Therefore, the~force exerted by the interface on the particle is
\begin{eqnarray} \label{eq:ch7_linear_force}
F_n = \left\{
	\begin{array}{ll}
k_s \left( \Delta z_n + R \cos\theta \right)  &  \text{for}~|\Delta z_n| \le R \\
0   & \text{otherwise}
	\end{array}
\right.
\end{eqnarray}
where $k_s$ is a spring constant, $\Delta z_n$ is the distance from the center of the particle to the instantaneous location of the liquid-vapor interface (which is flat for a film but curved for a droplet), and~$\theta$ is the contact angle of the solvent on the surface of the particle. Mathematically, $\Delta z_n = z_n - H(t)$, where $z_n$ is the particle's coordinate along the $z$-axis for a flat interface with the bottom of the film at $z=0$ or along the radial direction for a spherical interface with the center of the droplet at $z=0$. The~minimum of the confining potential is thus located at $H(t) - R \cos\theta$. In~Equation~(\ref{eq:ch7_linear_force}), a~positive (negative) value indicates that the force is toward the liquid solvent (vapor phase).

Some ambiguities exist in the literature regarding the physical interpretation of $k_s$. Pieranski proposed that $k_s = 2 \pi \gamma$ with $\gamma$ being the interfacial tension of the liquid-vapor interface~\cite{Pieranski1980}. However, this expression completely neglects capillary effects. Recently Tang and Cheng~\cite{Tang2018PRE} analyzed the capillary force exerted on a spherical particle of radius $R$ at a liquid-vapor interface when the particle is displaced out of its equilibrium location and showed that the linear approximation in Equation~(\ref{eq:ch7_linear_force}) generally works well but the spring constant should be understood as~\cite{Tang2018PRE, Joanny1984}
\begin{equation}\label{eq:ch7_spring_const}
k_s = \frac{2 \pi \gamma}{\ln (2L/R)}
\end{equation}
where $L$ is the lateral span of the liquid-vapor interface. In the simulations reported here, we set $k_s = 3.0\epsilon/\sigma^2$ unless otherwise noted. We also typically set $\theta = 0$, in~which case the potential barrier representing the liquid-vapor interface becomes the right half of a harmonic potential. Such a potential ensures that all the solutes are confined in the solution phase. The~contact angle $\theta$ can also be adjusted to tune the interaction between the solute and liquid-vapor interface. For~example, a~nonzero $\theta$ can be used for systems with attractive solute-interface interactions~\cite{Bigioni2006}.

\section{Applications of Moving Interface~Method} \label{sec:ch7_applications}
\unskip

\subsection{Drying of Solution Films of Polymer-Nanoparticle~Mixtures}

Evaporation of a mixed solution of polymer chains and nanoparticles has been studied via molecular dynamics (MD) simulations with both explicit~\cite{Cheng2016} and implicit~\cite{Howard2017} solvent models. Here we apply the moving interface method to study a mixture of 3200 linear polymer chains, each consisting of 100 connected beads of mass $m$ and diameter $\sigma$, and~a varying number of nanoparticles. All the beads interact through a Lennard-Jones (LJ) 12-6 potential, 
\begin{equation} \label{eq:ch7_lj}
U_{\text{LJ}}(r) = 4 \epsilon \left[ \left(\frac{\sigma}{r}\right)^{12} - \left(\frac{\sigma}{r}\right)^6   - \left(\frac{\sigma}{r_c}\right)^{12} + \left(\frac{\sigma}{r_c}\right)^6 \right]
\end{equation}
where $r$ is the separation of two beads, $\epsilon$ sets the strength of interaction, and~$r_c$ is the cutoff of the potential. To~model a good solvent, in~this study the LJ interactions are truncated at $r_c = 2^{1/6}\sigma$, rendering the potentials purely repulsive. Alternatively, one can increase the cutoff to include attractive interactions. However in this case the temperature of the solution must be above its theta temperature. Otherwise, phase separation will occur in the initial dilute~solution.

Adjacent beads on a chain are connected by a spring described by a finitely extensible nonlinear elastic (FENE) potential,
\begin{equation} \label{eq:ch7_fene}
U_{\text{FENE}}(r) = U_{\text{LJ}}(r) - \frac{1}{2} k R_o^2 \ln \left(1 - \frac{r^2}{R_o^2}\right)
\end{equation}
where $R_o = 1.5 \sigma$ and $k = 30 \epsilon/\sigma^2$~\cite{Kremer1990}. The LJ term in the FENE potential is truncated at $2^{1/6}\sigma$. With~these parameters, the~model describes polymer chains in a good solvent with a root-mean-square radius of gyration $R_g\simeq 7.2\sigma$ for 100-bead~chains.

A nanoparticle is modeled as a uniform sphere of LJ beads at a mass density of 1.0 m$/\sigma^3$~\cite{Everaers2003, IntVeld2008}. The interaction between a nanoparticle and a monomer bead on a polymer chain is given by an integrated LJ potential~\cite{Everaers2003},
\begin{align} \label{eq:ch7_pot_ns}
U_{\text{np}}(r) = & \frac{2}{9}\frac{R^3 \sigma^3 A_{\text{np}}}{\left(R^2 - r^2\right)^3} \nonumber \\
& \times \left[ 1 - \frac{\left(5R^6 + 45R^4 r^2 + 63 R^2 r^4 + 15 r^6\right)\sigma^6}
{15\left(R - r\right)^6 \left(R + r\right)^6} \right]
\end{align}
where $R$ is the nanoparticle radius, $r$ is the center-to-center distance between the nanoparticle and monomer, and~$A_{\text{np}}$ is a Hamaker constant setting the interaction strength. Here nanoparticles with two different radii, $R_m=10\sigma$ or $R_s=2.5\sigma$, are studied. To~facilitate the discussion below, we call the former medium nanoparticles (MNPs) and the latter small nanoparticles (SNPs). In~both cases, we set $A_{\text{np}} = 100\epsilon /\sigma^2$. The~nanoparticle-polymer interaction is purely repulsive with the potential in Equation~(\ref{eq:ch7_pot_ns}) truncated at $10.858\sigma$ for MNPs and $3.34\sigma$ for~SNPs.

The nanoparticle-nanoparticle interaction is given by an integrated LJ potential for two spheres~\cite{Everaers2003},
\begin{align} \label{eq:ch7_pot_nn}
U_{\text{nn}}(r) &= U_A(r) + U_R(r)
\end{align}
with
\begin{align} \label{eq:ch7_pot_ab}
U_A(r) = & -\frac{A_{\text{nn}}}{6} \left[ \frac{2 R_1 R_2}{r^2 - \left(R_1 + R_2\right)^2} + \frac{2 R_1 R_2}{r^2 - \left(R_1 - R_2\right)^2} \right. \nonumber \\
& \left. + \ln \left(  \frac{r^2 - \left(R_1 + R_2\right)^2}{r^2 - \left(R_1 - R_2\right)^2} \right) \right]  \\
U_R(r) &= \frac{A_{\text{nn}}}{37800}\frac{\sigma^6}{r}  \nonumber \\
& \times \left[ \frac{r^2 - 7r\left(R_1 + R_2\right) + 6 \left(R_1^2 + 7 R_1 R_2 + R_2^2\right)}{\left(r - R_1 - R_2\right)^7} \right. \nonumber \\
 &+ \frac{r^2 + 7r\left(R_1 + R_2\right) + 6 \left(R_1^2 + 7 R_1 R_2 + R_2^2\right)}{\left(r + R_1 + R_2\right)^7} \nonumber \\
 &- \frac{r^2 + 7r\left(R_1 - R_2\right) + 6 \left(R_1^2 - 7 R_1 R_2 + R_2^2\right)}{\left(r + R_1 - R_2\right)^7} \nonumber \\ 
 & \left. - \frac{r^2 - 7r\left(R_1 + R_2\right) + 6 \left(R_1^2 - 7 R_1 R_2 + R_2^2\right)}{\left(r - R_1 + R_2\right)^7} \right]
 \end{align}
where $R_1$ and $R_2$ are the radii of the nanoparticles, $r$ is the distance between their centers, and~$A_{\text{nn}}$ is a Hamaker constant. Here we use $A_{\text{nn}} = 39.48\epsilon$~\cite{Everaers2003}. The~nanoparticle-nanoparticle interaction is also purely repulsive with the potential truncated at $20.574\sigma$ for $R_1=R_2=10\sigma$ or $5.595\sigma$ for $R_1=R_2=2.5\sigma$. In~this manner, the~nanoparticles and polymer chains are guaranteed to be well dispersed in the implicit solvent prior to~evaporation.

The mixed solution of the nanoparticles, consisting of either 171 MNPs or 10944 SNPs, and~polymer chains is placed in a rectangular simulation cell with dimensions of $L_x \times L_y \times L_z$, where $L_x = L_y = 200 \sigma$ and $L_z = 800 \sigma$. Periodic boundary conditions are used in the $xy$ plane. Along the $z$ direction, the~solution is confined from below by a wall located at $z=0$. The~interaction between a particle (either a nanoparticle or a polymer bead) and the wall is given by a LJ 9-3 potential,
\begin{equation} \label{eq:lj_93}
U_{\text{w}}(h) = \epsilon_{\text{w}} \left[ \frac{2}{15}\left(\frac{a}{h}\right)^{9} - \left(\frac{a}{h}\right)^3   - \frac{2}{15}\left(\frac{a}{h_c}\right)^{9} + \left(\frac{a}{h_c}\right)^3 \right]
\end{equation}
with $\epsilon_{\text{w}} = 2.0\epsilon$, $a$ being the particle radius, $h$ as the shortest distance between the particle center and the wall. The~particle-wall potential is truncated at $h_c = 0.858a$ to make the wall purely repulsive. Prior to evaporation, the~solution is confined from above by the liquid-vapor interface, which is modeled as a potential barrier as in Equation~(\ref{eq:ch7_linear_force}) with $H(0) = 800 \sigma$. The~contact angle $\theta = 0$ is used for both polymer beads and~nanoparticles.

All the simulations reported here are performed with the Large-scale Atomic/Molecular Massively Parallel Simulator (LAMMPS)~\cite{Thompson2022}. A velocity-Verlet algorithm with a time step of $dt = 0.005\tau$ is used to integrate the equation of motion. All the particles in the system are thermalized at a temperature of $T = 1.0 \epsilon/k_B$ with a Langevin thermostat. The~damping time $\Gamma$ is set to $1\tau$ for the polymer beads, $6.25\tau$ for SNPs, and~$100\tau$ for MNPs, respectively. The~drying process is modeled by moving the liquid-vapor interface downward to $H_f = 200 \sigma$ at a constant speed $v_e$. Simulations are run for $v_e = 0.024 \sigma/\tau$ and $0.006 \sigma/\tau$.

After equilibration, each initial solution is a uniform mixture of the polymer chains and nanoparticles. The~initial volume fraction is $\phi_{n,0} = 0.0224$ for both SNPs and MNPs. The~initial number density of the polymer beads is $\rho_{p,0} = 0.01\sigma^{-3}$. The~size of a polymer chain can be quantified by its root-mean-square radius of gyration, $R_g$. The~100-bead chains have swollen conformations in the implicit solvent adopted here, similar to those in an explicit LJ solvent~\cite{Huang2021}, and $R_g \simeq 7.2\sigma$. Therefore, $R_m > R_g > R_s$. The~diffusion coefficients of the polymer chains and nanoparticles in each mixed solution are calculated with independent MD simulations. For~the polymer chains, the~diffusion coefficients, $D_p \simeq 0.00915\sigma^2/\tau$ in the mixtures with MNPs and $D_p \simeq 0.00949\sigma^2/\tau$ in the mixtures with SNPs. The~small difference in $D_p$ is likely caused by the different surface area of the nanoparticles in each mixture. At~the same volume fraction, the~total surface area of SNPs is 4 times that of MNPs. The~diffusion coefficient of MNPs is $D_m \simeq 0.0138\sigma^2/\tau$ while of SNPs is $D_s \simeq 0.0971\sigma^2/\tau$. Note that even though $R_m > R_g$, the~computed results show $D_m > D_p$, indicating that the diffusion coefficients do not follow the Stokes--Einstein relation.This is not surprising as in the implicit solvent adopted here, the~viscous damping is applied to each polymer bead. The~polymer dynamics thus follows the Rouse model instead of the Zimm model~\cite{RubinsteinColby2003} and a polymer chain freely drained by the implicit solvent cannot be treated as a solid object with a size equal to its $R_g$. With~the damping time of $\Gamma_p =1\tau$ for each bead, the~diffusion coefficient of a 100-bead chain is estimated to be $k_\text{B}T/(100\xi_p)=\Gamma_p k_\text{B}T/(100m)=0.01\sigma^2/\tau$, which is very close to the computed value of $D_p$. For~the nanoparticles, $D_s/D_m \simeq 7$, which is slightly larger than the value ($\simeq 5.8$) for the same sized nanoparticles in an explicit solvent~\cite{Tang2019JCP_compare}. Both are larger than the size ratio $R_m/R_s=4$. As~discussed in Section~\ref{ss:ch7_langevin}, the~diffusion coefficient of a particle following the Langevin dynamics is $D=\Gamma_n k_\text{B}T/m_n$ in the dilute limit, where $\Gamma_n$ is the damping time of the particle and $m_n$ is its mass. The~expected diffusion coefficient in the dilute limit for SNPs is $0.0955\sigma^2/\tau$, which is close to the calculated value in the polymer-SNP mixture. For~MNPs, the~diffusion coefficient in the dilute limit is expected to be $0.0239\sigma^2\tau$, which is $1.7$ times the computed value in the polymer-MNP~mixture.

Using the above results, the~P\'{e}clet number of each species can be computed as $\text{Pe}=H(0)v_e/D$, where $H(0)$ is the initial film thickness, $v_e$ is the receding speed of the liquid-vapor interface, and~$D$ is the diffusion coefficient. The~P\'{e}clet number thus characterizes the competition between evaporation-induced migration and particle diffusion. The~values of Pe for the polymer-nanoparticle mixtures studied here are summarized in Table~\ref{tb:system}, where $\text{Pe}_n$ is the P\'{e}clet number for the nanoparticles while $\text{Pe}_p$ is for the polymer~chains.

\begin{table}[ht]
\centering
\caption{P\'{e}clet numbers for all polymer-nanoparticle mixtures studied.}
\begin{tabular}{cccc}
\hline
System  & $v_e\tau/\sigma$ & $\text{Pe}_n$ & $\text{Pe}_p$  \\ \hline
MNP-1   & $0.024$ & 1391  & 2098 \\
MNP-2   & $0.006$ & 348   & 525 \\ 
SNP-1   & $0.024$ & 198   & 2023 \\
SNP-2   & $0.006$ & 49    & 506 \\
\hline
\end{tabular}
\label{tb:system}
\end{table}

\begin{figure}[hbt]
  \includegraphics[width = 0.5\textwidth]{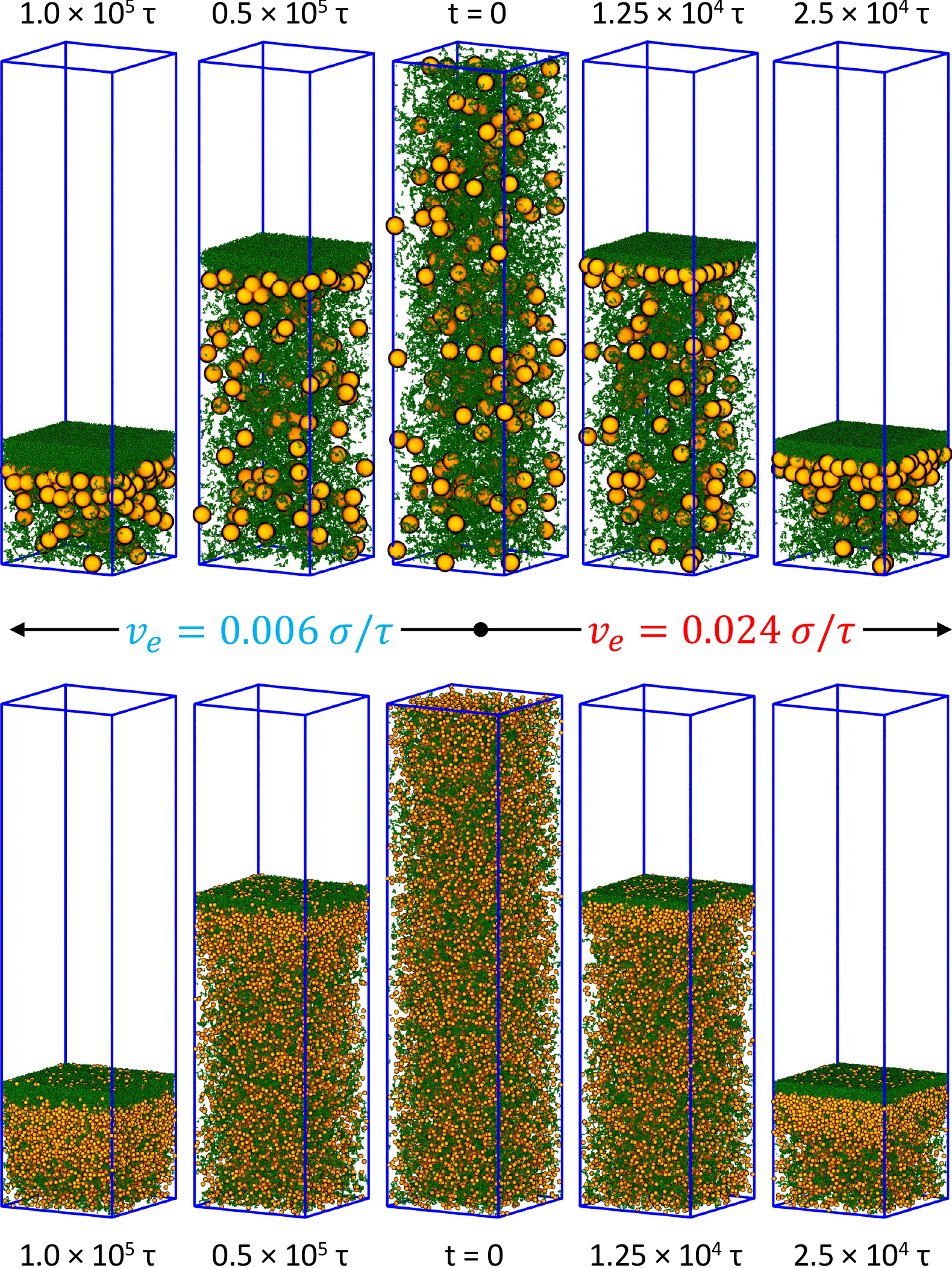}
    \caption{Snapshots of solutions of nanoparticles (orange) and polymer chains (green) at various stages of drying for different evaporation rates: (\textbf{left}) $v_e = 0.006\sigma/\tau$ and (\textbf{right}) $v_e=0.024 \sigma/\tau$; (\textbf{top}) the polymer-MNP mixtures and (\textbf{bottom}) the polymer-SNP~mixtures.}
  	\label{fg:ch7_pnc_snapshots}
\end{figure}

\begin{figure*}[hbt]
 \centering 
  \includegraphics[width = 1.0\textwidth]{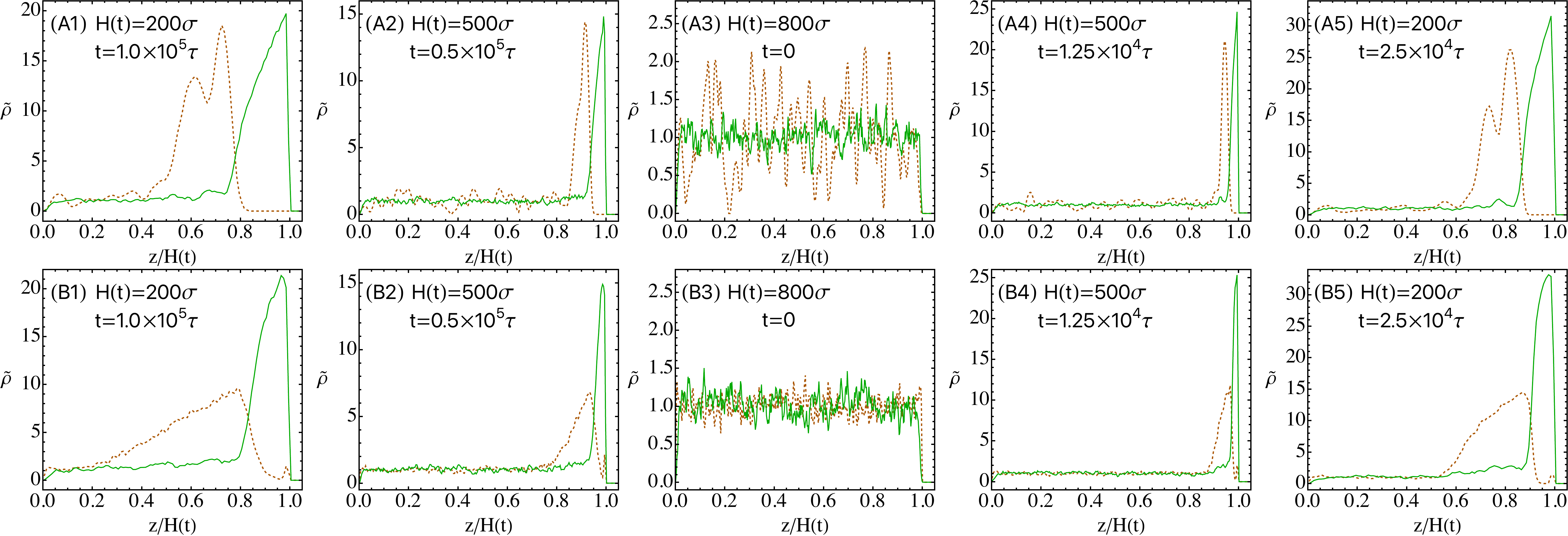}
  \caption{Density profiles of nanoparticles (orange dashed lines) and polymer chains (green solid lines) at various stages of drying for the polymer-MNP mixtures (\textbf{A1-A5}
) and the polymer-SNP mixtures (\textbf{B1-B5}). The~profiles have a one-to-one correspondence to the snapshots in Figure~\ref{fg:ch7_pnc_snapshots}. The~equilibrium density profiles are shown in A3 and B3. The~sequences A3$\rightarrow$A2$\rightarrow$A1 and B3$\rightarrow$B2$\rightarrow$B1 are for $v_e = 0.006 \sigma/\tau$, while the sequences A3$\rightarrow$A4$\rightarrow$A5 and B3$\rightarrow$B4$\rightarrow$B5 are for $v_e = 0.024 \sigma/\tau$.}
  	\label{fg:ch7_pnc_den_profile}
\end{figure*}

The snapshots of the mixed solutions undergoing drying are shown in Figure~\ref{fg:ch7_pnc_snapshots} for various times during evaporation and the corresponding density profiles of the nanoparticles and polymer chains are shown in Figure~\ref{fg:ch7_pnc_den_profile}. Prior to evaporation, all the chains and nanoparticles are uniformly distributed in the solution. After~drying, the~polymer is enriched near the descending interface for all four systems studied, which feature two nanoparticle sizes and two evaporation rates. Because~of the repulsion with the polymer chains, the~nanoparticles are expelled from this skin layer of polymer and become accumulated just below the skin layer. As~a result, a~stratified state is created, similar to the results previously obtained with an explicit solvent model~\cite{Cheng2016}. In the implicit solvent simulation of Howard~et~al., similar stratification is also observed with nanoparticles concentrated below a concentrated layer of polymer chains~\cite{Howard2017b}. However, a~layer of nanoparticles is also found on top of the polymer skin layer in those simulations (see Figures~6 and 7 of Ref.~\cite{Howard2017b}). This is an effect of the potential barrier used in Ref.~\cite{Howard2017b} to represent the liquid-vapor interface, where the contact angle was set to $90^\circ$ for both nanoparticles and polymer chains. The~resulting potential is thus attractive for both species and the attraction can be quite strong for nanoparticles, leading to their adsorption at the interface. Since here we use $\theta = 0$ for all the solutes, this effect is removed, as~evident in Figure~\ref{fg:ch7_pnc_snapshots} (see the results for the polymer-MNP mixtures).

Although $R_m > R_g$, the~results in Figure~\ref{fg:ch7_pnc_snapshots} for the polymer-MNP mixtures cannot be classified as ``small-on-top'' stratification~\cite{Fortini2016,Fortini2017,Zhou2017,Howard2017,Tang2018Langmuir} as the polymer chains actually diffuse more slowly than the nanoparticles. In~each mixed solution, the~polymer chains have a larger P\'{e}clet number and should be effectively treated as the ``larger'' species. However, they are always found on top as a skin layer after drying. When the evaporation rate is reduced, the~skin layer becomes thicker, signaling enhanced stratification. Such ``polymer-on-top'' stratification is also found in mixtures with SNPs that have a much smaller radius (see the bottom row of Figure~\ref{fg:ch7_pnc_snapshots}) and in other reports~\cite{Howard2017b}. A~question naturally arises: Does ``polymer-on-top'' stratification always occur in all polymer-particle mixtures that undergo a drying process? Or equivalently, can ``particle-on-top'' stratification be realized in drying polymer-particle mixtures? Although these questions are still open at this point, a~careful examination of the results shown in Figure~\ref{fg:ch7_pnc_snapshots} has offered certain clues. In~particular, for~the polymer-SNP mixtures some SNPs are observed on-top-of the polymer skin layer during drying. Since $R_g/R_s \simeq 3$ and $D_p/D_s\simeq 0.1$, the~results indicate that the polymer-SNP mixtures may have a tendency of exhibiting ``small-on-top'' stratification (in this case, ``particle-on-top'') but are not there yet as the size ratio of $3$ is still relatively small. Furthermore, when the evaporation rate is reduced by a factor of 4 from $v_e = 0.024\sigma/\tau$ to $0.006\sigma/\tau$, the~number of SNPs above the polymer skin layer slightly increases, though~the skin layer thickens in this case. These results indicate that to realize ``particle-on-top'' stratification, a~even larger ratio between the chain size and the nanoparticle radius may be needed, which can be achieved by using longer chains. A~slower evaporation rate may further favor ``particle-on-top'' stratification. Work along this line will be reported in the~future.

The distribution of the nanoparticles and polymer chains shown in Figure~\ref{fg:ch7_pnc_snapshots} can be quantified by their density profiles along the direction of drying, as~shown in Figure~\ref{fg:ch7_pnc_den_profile}. The~local number density of polymer beads is defined as $\rho_p(z)=n_p(z)/\left(L_xL_y\Delta z\right)$, where $n_p(z)$ is the number of polymer beads with their $z$-coordinates between $z-\Delta z /2$ and $z+\Delta z /2$. The~local mass density of nanoparticles is defined as $\rho_n(z)=n_n(z) m_n /\left( L_xL_y\Delta z \right)$, where $n_n(z)$ is the partitioned contribution to the slice parallel to the $xy$ plane and from $z-\Delta z /2$ to $z+\Delta z /2$ from all the nanoparticles straddling that slice, weighted by their intersection volume with the slice, and~$m_n$ is the nanoparticle mass. For~the results shown in Figure~\ref{fg:ch7_pnc_den_profile}, $\Delta z =2 \sigma$. To~facilitate comparison, the~local number density of polymer beads is normalized by their initial average number density ($0.01\sigma^{-3}$) and the local mass density of nanoparticles is normalized by their initial average mass density ($0.0224m\sigma^{-3}$). The~normalized density profiles ($\Tilde{\rho}$) in Figure~\ref{fg:ch7_pnc_den_profile} reveal several features hidden in the snapshots. First of all, as~the evaporation rate is reduced, the~peak height of the density profile decreases for both nanoparticles and chains but the range where the density profile exhibits gradients is widened. This indicates that the perturbation caused by the receding interface propagates farther as it take longer time to achieve a certain stage of drying at slower evaporation rates. Secondly, at~the same evaporation rate and stage of drying, the~peak height as well as the gradient range and magnitude of the polymer density are similar in the two mixtures containing nanoparticles with different diameters but the larger nanoparticles develop a higher peak in their density profile. Furthermore, the~gradient of the density profile occurs in a smaller spatial range and as a result, the~corresponding density profile has a steeper gradient for the larger nanoparticles. The~density profiles in Figure~\ref{fg:ch7_pnc_den_profile} and the snapshots in Figure~\ref{fg:ch7_pnc_snapshots} therefore both point to a more dramatic ``polymer-on-top'' stratification when the nanoparticles are larger than the polymer chains. Finally as shown in the bottom row of Figure~\ref{fg:ch7_pnc_den_profile}, there is an excess of SNPs near the liquid-vapor interface in the equilibrium solution as they are smaller than the polymer chains and can get closer to the interface~\cite{Tang2018Langmuir}, which leads to a weak peak in the density profile of SNPs above the polymer skin layer during evaporation. The~presence of such SNPs near the receding interface is also visible in the snapshots shown in the bottom row of Figure~\ref{fg:ch7_pnc_snapshots}.

\subsection{Drying of Suspension Droplets of Bidisperse Mixtures of~Nanoparticles}
\label{sec:np_droplet}

\begin{figure*}[htb]
  \includegraphics[width = 0.7 \textwidth]{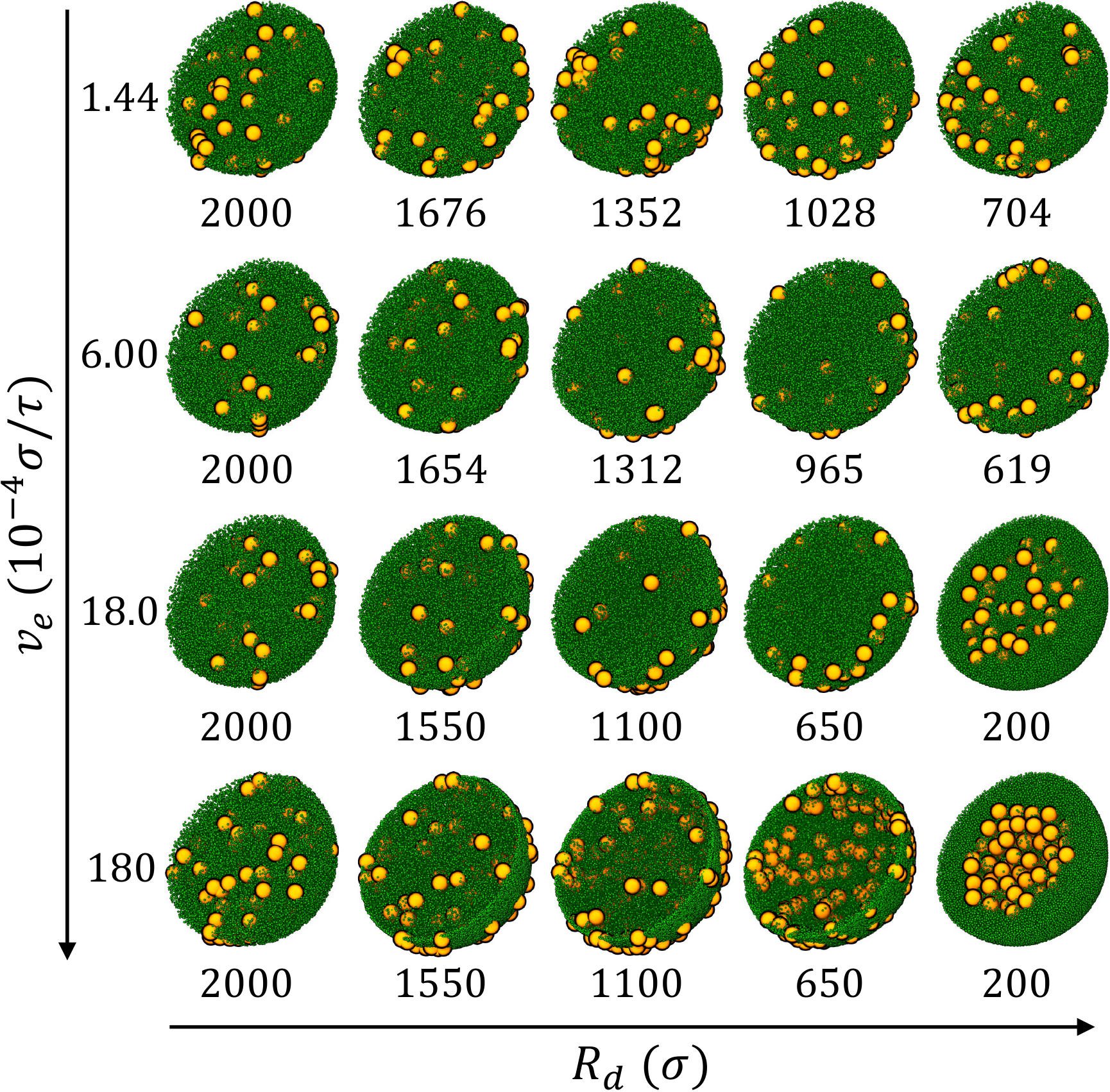}
    \caption{Snapshots of drying droplets of bidisperse mixtures of LNPs (orange) and SNPs (green) at various evaporation rates ($v_e$). The~droplet radius ($R_d$) is indicated below each snapshot. All the droplets are sliced through the center to show the interior and scaled to the same size to improve visualization.}
  	\label{fg:ch7_codrop_snapshots}
\end{figure*}

Stratification has mostly been discussed in the context of drying films~\cite{Fortini2016, Fortini2017, Makepeace2017, Howard2017, Howard2017b, Tatsumi2018, Statt2018, Tang2018Langmuir}. However, stratification can also occur in drying droplets under appropriate conditions~\cite{Liu2019, Xiao2019, Gartner2020, Howard2020}, which may lead to fast procedures of making core-shell structures. In~industry, spray drying processes are frequently practiced, where droplet drying is a critical step~\cite{Lintingre2016}. In one such process, a~particle suspension is injected from a nozzle or an injector into a flowing gas. The~liquid jet of the suspension then breaks into droplets, which further dry in the hot gas into clusters of particles (i.e., solutes). The~drying of a single droplet was recently studied using the Leidenfrost effect: a droplet is levitated on a hot substrate by its own vapor and then let dry~\cite{Lintingre2015}. Here we use the moving interface method to study the drying of a suspension droplet of a mixture of nanoparticles of two different radii, motivated by the possibility of creating a core-shell cluster with one type of nanoparticles in the outside shell while the other type in the inner core. In~this context, a~bidisperse nanoparticle mixture stratifies radially into either ``small-on-outside'' or ``large-on-outside''.

The droplet contains 200 large nanoparticle (LNPs) of radius $R_l = 20\sigma$ and 102400 SNPs of radius $R_s = 2.5\sigma$, which are initially confined by a spherical potential barrier inside a sphere with radius $H(0)=2000\sigma$. Their initial volume fractions are $\phi_l = \phi_s = 0.0002$. Although~the liquid-vapor interface of a droplet is curved, we still adopt Equation~(\ref{eq:ch7_linear_force}) for the particle-interface interaction~\cite{Lishchuk2016}. We also set the contact angle $\theta = 0$ for both LNPs and SNPs. The~evaporation process is mimicked by decreasing the radius of the droplet at a constant speed, $v_e$. The~instantaneous radius at time $t$ since the start of the evaporation is thus $R_d = H(0) - v_e t$. The~nanoparticle-nanoparticle interactions are given by Equation~(\ref{eq:ch7_pot_nn}) with $A_\text{nn} = 39.48\epsilon$. All these interactions are purely repulsive with the corresponding potentials truncated at $40.571\sigma$, $23.086 \sigma$, and~$5.595 \sigma$ for the LNP-LNP, LNP-SNP, and~SNP-SNP pairs, respectively. A~Langevin thermostat is used to maintain the temperature of the system at $T = 1.0 \epsilon/k_B$. To~conserve the Stokes--Einstein relation, the~damping time is set to $1600 \tau$ for LNPs and $25\tau$ for SNPs according to Equation~(\ref{eq:ch7_damping}). Four independent drying simulations are performed with $v_e$ ranging from $1.8\times 10^{-2} \sigma/\tau$ to $1.44\times 10^{-4} \sigma/\tau$.

\begin{figure*}[htb]
  \includegraphics[width = 0.7 \textwidth]{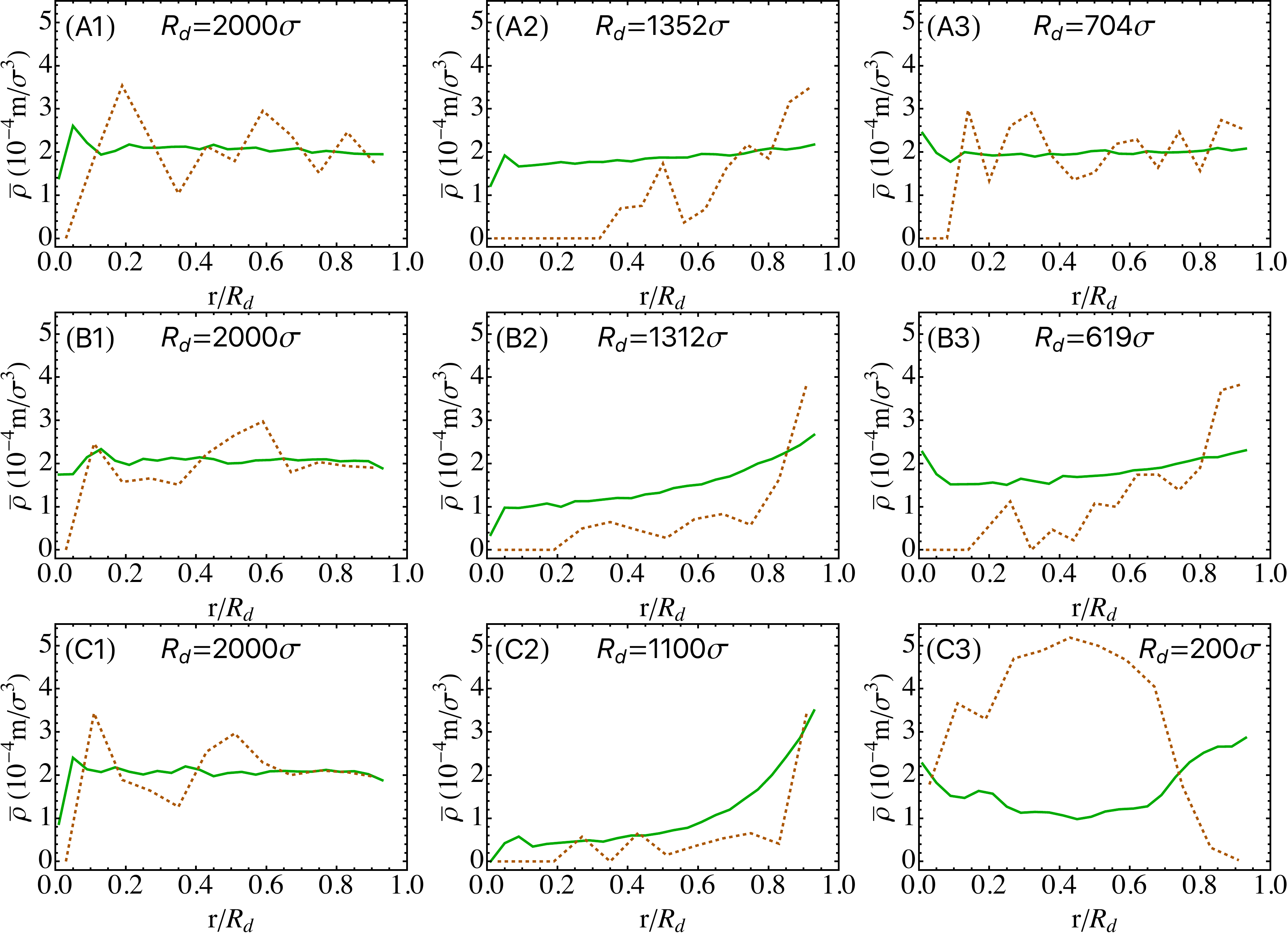}
    \caption{Density profiles of SNPs (solid green lines) and LNPs (dashed brown lines) in the droplet along its radial direction at various evaporation rates: $v_e = 1.44\times 10^{-4}\sigma/\tau$ (\textbf{A1-A3}), $v_e = 6\times 10^{-4}\sigma/\tau$ (\textbf{B1-B3}), and~$v_e = 1.8\times 10^{-3}\sigma/\tau$ (\textbf{C1-C3}). The~stage of drying is indicated by the droplet radius ($R_d$).}
  	\label{fg:ch7_codrop_den_profile}
\end{figure*}

The snapshots of the droplets at various stages of drying are shown in Figure~\ref{fg:ch7_codrop_snapshots} and the density profiles of nanoparticles are shown in Figure~\ref{fg:ch7_codrop_den_profile}. The~density of nanoparticles is defined as $\rho(r)=n_im_i/\left[V(r+\Delta r)-V(r)\right]$, where $n_i$ is the number of $i$-type nanoparticles in a spherical shell from $r$ to $r+\Delta r$, $m_i$ is the mass of one $i$-type nanoparticle, and~$V(r)=\frac{4}{3}\pi r^3$ is the volume of a sphere of radius $r$. For~a nanoparticle straddling multiple shells, its mass is partitioned to each shell in proportion to its intersection volume with that shell. During~evaporation, the~droplet radius $R_d$ decreases and the average density of nanoparticles increases as drying progresses. To~facilitate comparison, the~local density $\rho(r)$ is scaled as $\overline{\rho}(r) = \rho(r)/\beta^3$, where $\beta = H(0)/R_d$. The~scaled density profiles are plotted in Figure~\ref{fg:ch7_codrop_den_profile}.

After equilibration and prior to evaporation, the~droplet is a uniform mixture of LNPs and SNPs (see the first column of Figure~\ref{fg:ch7_codrop_snapshots}) with an average total nanoparticle density of $4\times 10^{-4}$~m/$\sigma^3$. The~diffusion coefficients are found to be $D_l = 0.0414 \sigma^2/\tau$ for LNPs and $D_s = 0.378 \sigma^2/\tau^2$ for SNPs with independent MD simulations. The~ratio of $D_s/D_l$ is about $9.1$, which is just slightly larger than the size ratio $R_l/R_s = 8$. Since the initial volume fraction of the nanoparticles in the droplet is very low, the~computed result of the diffusion coefficient is very close to its value in the dilute limit, which is $0.382\sigma^2/\tau$ for SNPs and $0.0477\sigma^2/\tau$ for LNPs, respectively. With~$D_l$ and $D_s$, the~P\'{e}clet number is then computed for both LNPs and SNPs at a given evaporation rate ($v_e$) and the results are included in Table~\ref{tb:Pe_droplet}.

\begin{table}[ht]
\centering
\caption{P\'{e}clet numbers for LNPs and SNPs in their mixtures.}
\begin{tabular}{ccc}
\hline
$v_e\tau/\sigma$ & $\text{Pe}_l$ & $\text{Pe}_s$  \\ \hline
$1.44\times 10^{-4}$ & 7.0    & 0.76 \\
$6.0\times 10^{-4}$  & 29   & 3.2 \\ 
$1.8\times 10^{-3}$  & 87   & 9.5 \\
$1.8\times 10^{-2}$  & 870  & 95 \\
\hline
\end{tabular}
\label{tb:Pe_droplet}
\end{table}

During evaporation, both SNPs and LNPs are first enriched near the liquid-vapor interface, though~the degree of enrichment is lower at slower drying rates. At~high drying rates (e.g., $v_e=1.8\times 10^{-2}\sigma/\tau$ and $1.8\times 10^{-3}\sigma/\tau$), because~of the similar physical mechanism leading to ``small-on-top'' stratification in drying films of bidisperse particles~\cite{Fortini2016, Zhou2017, Tang2018Langmuir}, SNPs form a concentrated shell near the interface while LNPs are pushed out of this region and into the interior of the droplet. In~the final state, a~``small-on-outside'' cluster is clearly observed (see the third and fourth row of Figure~\ref{fg:ch7_codrop_snapshots} and the bottom row of Figure~\ref{fg:ch7_codrop_den_profile}). The simple model discussed here thus points to the possibility of creating core-shell clusters of particles by drying suspension droplets rapidly. Real spray drying processes are of course more complicated with many factors that are not captured by our simple model, such as air invasion, cracking, and~buckling~\cite{Lintingre2015, Kooij2016}. Despite these limitations, our results indicate that increasing drying rates may lead to new strategies of controlling the structure of the resulting clusters or creating new~structures.

\begin{figure*}[htb]
  \includegraphics[width = 0.8 \textwidth]{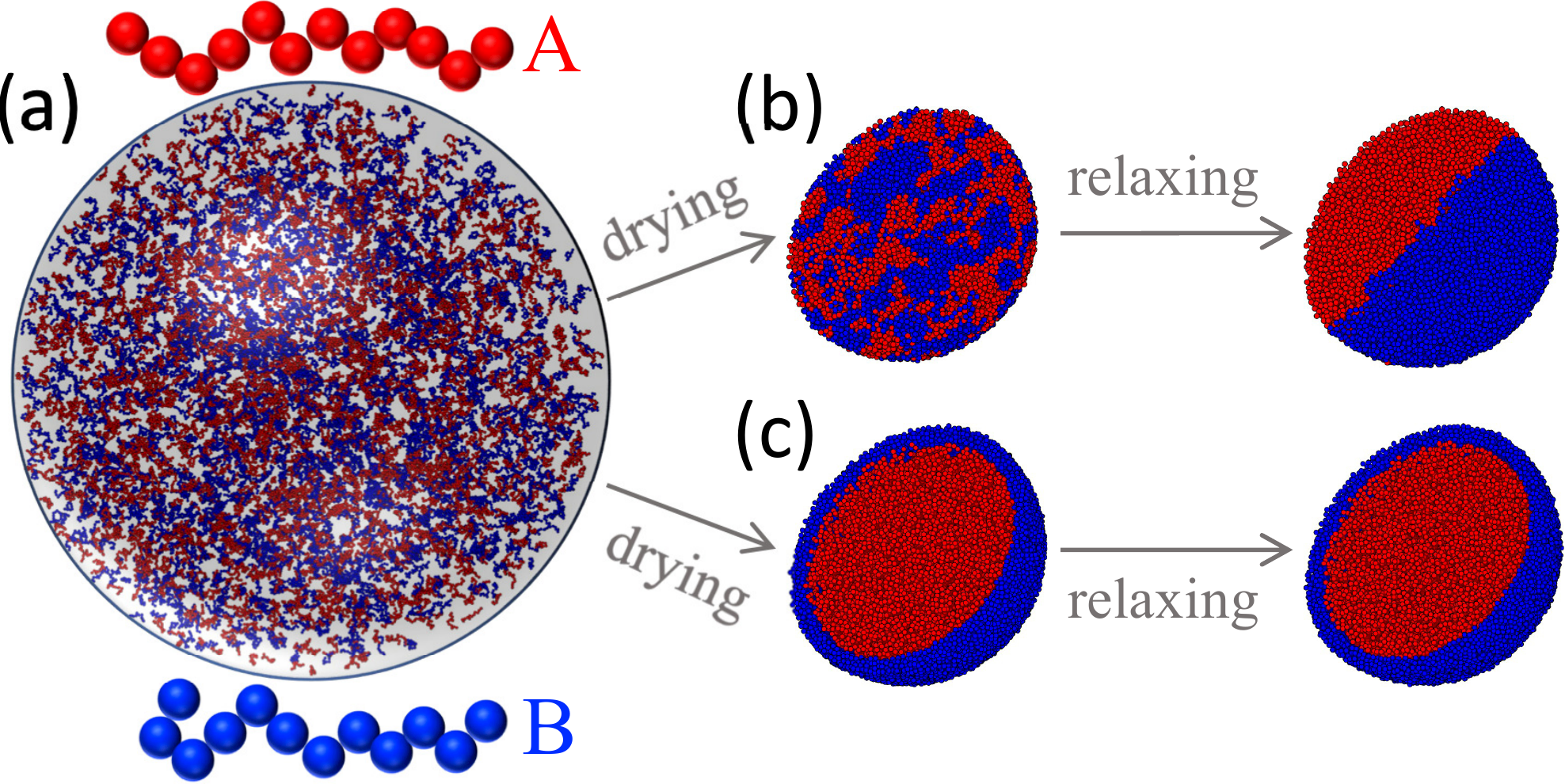}
    \caption{(\textbf{a}) A solution droplet of a polymer blend prior to drying. Depending on the interaction with the receding liquid-vapor interface, polymeric particle with different morphologies are obtained after drying: (\textbf{b}) A Janus particle is produced with $\theta = 0$ for both polymer A (red) and B (blue) after drying for $5 \times 10^3 \tau$ followed by relaxation for $2 \times 10^5 \tau$; (\textbf{c}) A core-shell particle is produced with $\theta = 0$ for polymer A (red) while $\theta = \pi / 2$ for polymer B (blue) after drying for $5 \times 10^3 \tau$ followed by relaxation for $5 \times 10^4 \tau$. For~clarity, the~droplets after drying and relaxation are sliced through the center to show the~interior.}
  	\label{fg:ch7_homopolymers}
\end{figure*}

When the drying rate is reduced by lowering $v_e$ to $6\times 10^{-4}\sigma/\tau$ (with $\text{Pe}_l = 29$ and $\text{Pe}_s = 3.2$), ``large-on-outside'' stratification is observed, as~shown in the second rows of Figure~\ref{fg:ch7_codrop_snapshots} and Figure~\ref{fg:ch7_codrop_den_profile}. During~drying, both SNPs and LNPs are accumulated at the receding liquid-vapor interface but the enrichment of LNPs is more significant. Eventually, LNPs are more enriched at the surface of droplet. When $v_e$ is further reduced to $1.44\times 10^{-4}\sigma/\tau$ (with $\text{Pe}_l = 7$ and $\text{Pe}_l = 0.76$), a~uniform distribution of LNPs and SNPs is found in the final dried droplet (see the first rows of Figure~\ref{fg:ch7_codrop_snapshots} and Figure~\ref{fg:ch7_codrop_den_profile}). In~this case, evaporation is slow and SNPs are almost uniformly distributed in the droplet in the entire process of drying. LNPs are first accumulated near the receding interface in the early stage of drying but eventually become uniformly distributed too. The~simulation results thus demonstrate that the distribution of nanoparticles transitions from uniform to ``large-on-outside'' to ``small-on-outside'' as the evaporation rate is increased, which is consistent with the prediction of the diffusive model of stratification in drying films proposed by Zhou~et~al.~\cite{Zhou2017}.

\subsection{Drying of Solution Droplets of a Polymer~Blend}
\label{sec:blend}

The moving interface method can also be used to study the drying process of polymer solution droplets. In~this section we focus on the solution of a polymer blend, which is a symmetric mixture of short polymer chains of type A and B, with~2048 chains of each type. Each chain is linear and consists of 12 beads of mass $m$ that are bonded by the FENE potential in Equation~(\ref{eq:ch7_fene})~\cite{Kremer1990}. In addition to the bonded interaction, all the nonbonded pairs of beads interact via a LJ potential with a cutoff distance of $2^{1/6} \sigma$, i.e.,~a purely repulsive potential. The~systems are kept at $T=1.0\epsilon/k_\text{B}$ via a Langevin thermostat with a damping time of $\Gamma = 10\tau$. As~a result, at~low concentrations the polymer chains adopt swollen conformations as in a good solvent. To~model an incompatible polymer blend, the~strength of the self interaction is set to $\epsilon_{AA} = \epsilon_{BB} = \epsilon$ while the cross repulsion is stronger with $\epsilon_{AB} = 2.0 \epsilon$. The~critical value of $\epsilon_{AB}$ for a symmetric blend to phase separate depends on its density. For~a melt of a symmetric mixture of linear $N$-bead chains at a density of $0.85m/\sigma^3$, Grest~et~al. found that it phase separates into an ordered phase at $\epsilon_{AB} \gtrsim \epsilon\left( 1 + 3.4/N \right)$~\cite{Grest1996}. Therefore, the~mixtures studied here are expected to phase separate as the packing density of the systems approaches the melt~density.

In the initial state prior to drying, all the chains are uniformly dispersed in a sphere with radius $100\sigma$, as~shown in Figure~\ref{fg:ch7_homopolymers}a. All the polymer beads are confined in this sphere with a spherical potential barrier as described in Section~\ref{sec:np_droplet} with $H(0) = 100\sigma$. The~potential given in Equation~(\ref{eq:ch7_linear_force}) with $k_s=3.0\epsilon/\sigma^2$ and $R=2\sigma$ is used for the interaction between a polymer bead and the liquid-vapor interface. The~initial density of each droplet is about $0.012m/\sigma^3$. During~drying, the~radius of the droplet is reduced to $H_f = 24 \sigma$ at a rate of $v_e = 1.52\times 10^{-2} \sigma/\tau$. At~this final radius, the~packing density of the polymer beads is about $0.85m/\sigma^3$, which is the melt density of the blend under an external pressure of about $5\epsilon/\sigma^3$~\cite{Grest1996}. The entire drying process thus lasts $5000\tau$.

Figure~\ref{fg:ch7_homopolymers} shows that the morphology of the resulting polymeric particle depends on the drying conditions and the polymer-interface interaction controlled by the contact angle $\theta$ in Equation~(\ref{eq:ch7_linear_force}). In~Figure~\ref{fg:ch7_homopolymers}b, $\theta = 0$ for both polymers. The~two polymers phase separate after fast drying as the cross repulsion (A-B) is sufficiently stronger than the self repulsion (A-A and B-B) in the two components~\cite{Grest1996}. Domains of each component are clearly visible in the polymeric particle. After~relaxation, a~Janus particle is produced with each component occupying half a~sphere.

\begin{figure*}[htp]
  \includegraphics[width = 0.8 \textwidth]{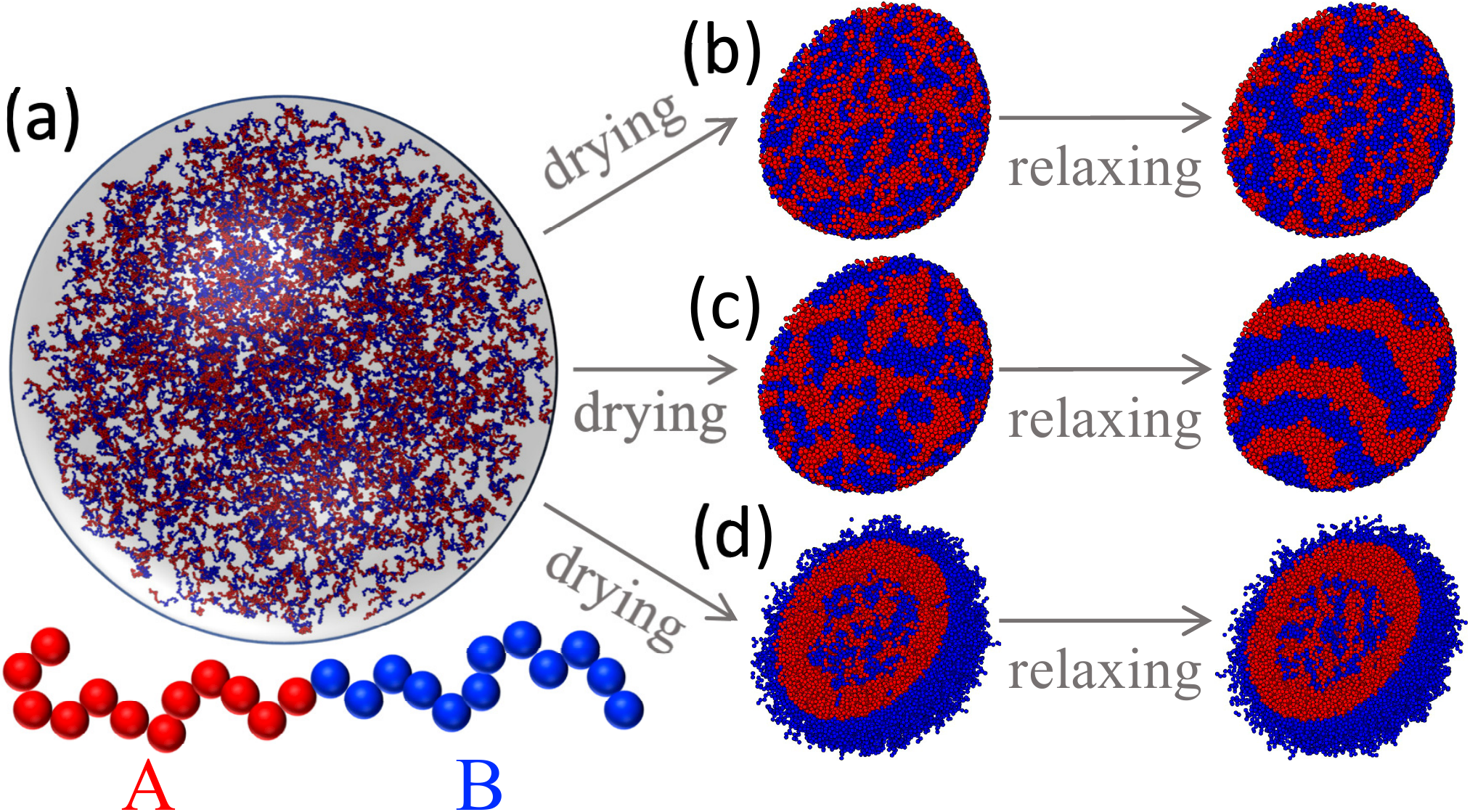}
    \caption{(\textbf{a}) A solution droplet of a diblock copolymer with $N=24$ prior to drying. Depending on the cross repulsion between blocks and the block-interface interaction, polymeric particles with different morphologies are obtained after drying: (\textbf{b}) A particle with small domains of A (red) and B (blue) blocks is produced with $\theta = 0$ for both blocks and $\epsilon_{AB}=2.0\epsilon$; (\textbf{c}) A particle with stripes of A and B blocks is produced with $\theta = 0$ for both blocks and $\epsilon_{AB}=8.0\epsilon$; (\textbf{d}) A particle with a mixture core enclosed by a bilayer shell (i.e., a~layer of B blocks on the outside and a layer of A blocks on the inside) is produced with $\theta = 0$ for block A (red), $\theta = \pi / 2$ for block B (blue), and~$\epsilon_{AB}=8.0\epsilon$. In~each case, the~droplet is dried to a radius of $24\sigma$ during $5 \times 10^3 \tau$, followed by relaxation for $2 \times 10^5 \tau$. For~clarity, the~droplets in (\textbf{b}--\textbf{d}) are sliced through the center to show the~interior.}
  	\label{fg:ch7_copolymers}
\end{figure*}

In Figure~\ref{fg:ch7_homopolymers}c, $\theta = 0$ for polymer A while $\theta = \pi/2$ for polymer B, indicating that the liquid-vapor interface is repulsive for polymer A but attractive for polymer B. The latter thus has a tendency to be adsorbed at the liquid-vapor interface. As~a result, the~B-type chains are enriched at the droplet surface during drying while all the A-type chains are pushed into the interior of the sphere. In~the final state, with~a radius of $24 \sigma$, a~core-shell distribution can be clearly identified. The~core-shell structure remains completely stable during the subsequent relaxation period. The~results thus indicate that the solute-interface interaction is an important factor affecting the structures produced by drying~\cite{Bigioni2006} and can be used to produce polymeric particles with different~morphologies.

\subsection{Drying of Solution Droplets of Diblock~Copolymers}

Block copolymers can also be employed to produce structured polymeric nanoparticles~\cite{Higuchi2008}. For a bulk system of a block copolymer with incompatible blocks, there are well-known ordered structures such as lamellar, cylindrical, and~spherical phases depending on the fraction of the two components on the chain~\cite{Bates1990}. In this section, we use a setup almost identical to the one in Section~\ref{sec:blend} to study the drying of a solution droplet of symmetric diblock copolymer chains. Here, the~number ratio between monomers in the A blocks and those in the B blocks is 1:1 and we vary the chain length $N$ from 12 to 96. The~total number of monomers is fixed at 49152 and the number of chains is thus $49152/N$. The~bonded interaction is given by the FENE potential in Equation~(\ref{eq:ch7_fene}). All the monomers, if~not directly bonded, interact with each other via a LJ 12-6 potential as in Equation~(\ref{eq:ch7_lj}). The~nonbonded interaction is purely repulsive. The~interaction strength in the LJ potential is $\epsilon$ for the A-A and B-B pairs. The~strength of A-B interactions is given by $\epsilon_{AB}$, which is varied from $2\epsilon$ to $8\epsilon$. A~Langevin thermostat with a damping time of $\Gamma = 10 \tau$ is used to control the temperature at $1.0\epsilon/k_\text{B}$. In~the initial state of each solution (e.g., see Figure~\ref{fg:ch7_copolymers}a for $N=24$), all the polymer chains are confined in a sphere with radius $100\sigma$ by a spherical potential barrier described by Equation~(\ref{eq:ch7_linear_force}) with $k_s = 3.0 \epsilon/\sigma^2$ and $R=1.0\sigma$. During~a drying period that lasts $5000\tau$, the~radius of the droplet is reduced to $H_f = 24\sigma$, yielding a packing density of $0.85m/\sigma^{3}$ for the polymer beads in the final~droplet.

Similar to the case of a polymer blend, the~results in Figure~\ref{fg:ch7_copolymers} for $N=24$ show that the structure of the resulting polymeric particle of diblock chains depends on the block-interface interaction and the strength of the block-block repulsion. The~polymer-interface interaction is determined by the contact angle $\theta$ in Equation~(\ref{eq:ch7_linear_force}). In~Figures~\ref{fg:ch7_copolymers}b,c, $\theta = 0$ for both blocks and the interface therefore appears neutral for all the monomers. The~value of $\epsilon_{AB}$ at which the order-disorder transition (ODT) occurs depends on the total density of the system. Grest~et~al.~\cite{Grest1996} found that for a symmetric diblock with $N=20$ at a density of $0.85m/\sigma^{3}$, ODT occurs at $\epsilon_{AB} \gtrsim 3.9\epsilon$. In~the current systems with $N=24$, the~critical value of $\epsilon_{AB}$ is expected to be slightly smaller. As~shown in Figure~\ref{fg:ch7_copolymers}b, with~$\epsilon_{AB}=2.0\epsilon$, only small domains of each type of blocks are observed after drying. No significant growth of the domain size is observed after relaxation. When the cross repulsion between the two blocks is increased to $\epsilon_{AB}=8.0\epsilon$, bringing the system deeply into the ordered-phase region in the phase diagram~\cite{Bates1990}, the blocks start to aggregate during drying, as shown in Figure~\ref{fg:ch7_copolymers}c. After relaxation, stripes of different blocks are clearly visible because of the incompatibility of the blocks, resembling the lamellar phase in bulk diblock copolymers~\cite{Bates1990}.

In Figure~\ref{fg:ch7_copolymers}d, nonneutral block-interface interactions are adopted with $\theta = 0$ for block A while $\theta = \pi/2$ for block B. As~a result, monomers in the B blocks have a tendency to adsorb at the liquid-vapor interface while the A blocks are repelled by the interface. After~drying, the~resulting particle is enveloped by a layer of B blocks adsorbed at the liquid-vapor interface, with~a layer of A blocks inside that is bonded to the B blocks at the surface. The~core region of the final particle is filled with a mixture of diblock chains that tend to phase separate. In~a larger drop filled with longer diblock chains, an~onion-like particle with layers of different blocks alternating radially is expected to form. Studies on this and other interesting morphologies will be reported in the~future.

\begin{figure}[htp]
  \includegraphics[width = 0.45 \textwidth]{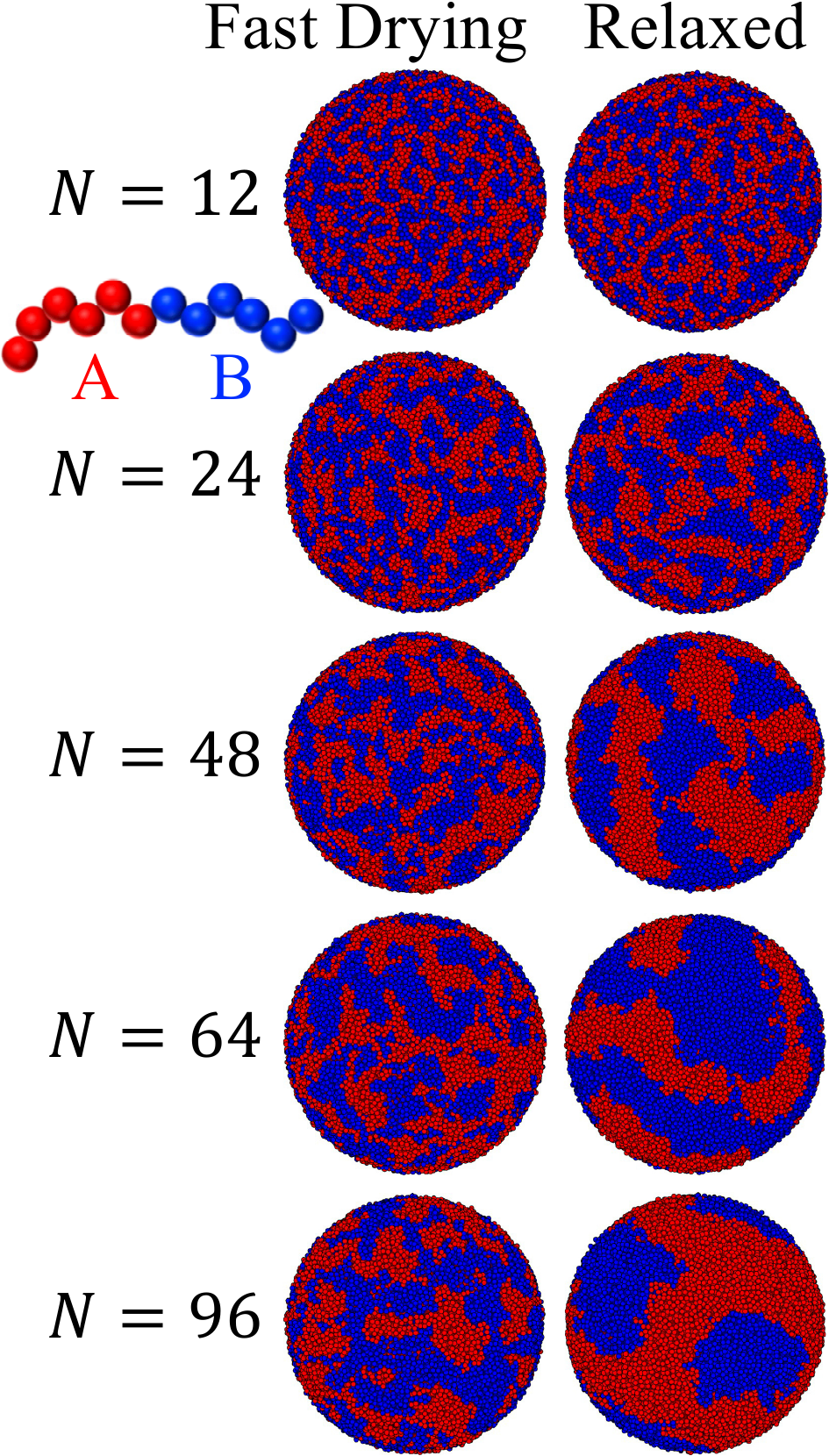}
    \caption{Effect of chain length ($N$) on particle morphology in drying droplets of diblock copolymer chains. For~all the systems, $\epsilon_{AB}=2.0\epsilon$ and $\theta = 0$ for both block A (red) and B (blue). The~drying process lasts $5 \times 10^3 \tau$, leading to particles in the top row with a radius of $24\sigma$. Relaxing each system for an additional period of $2 \times 10^5 \tau$ yields the particles in the bottom row.}
  	\label{fg:ch7_BCP_chain_length}
\end{figure}

Figure~\ref{fg:ch7_BCP_chain_length} shows the effect of chain length ($N$) on the particle morphology, while the length ratio of the two blocks is kept at 1:1. For~all these systems, $\epsilon_{AB}=2.0\epsilon$ and $\theta = 0$ for both blocks, indicating that the interface is neutral and repulsive for both components. After~fast drying to $H_f = 24\sigma$, the~resulting polymeric particle has a disordered structure at $N=12$ but phase separation occurs for longer chains, resulting in patchy polymeric articles. After~a relaxation process in which the radius of the confining spherical potential is fixed, rough stripe-like domains are formed in the polymeric particles with each domain dominated by one type of blocks. Each domain can be regarded as a patch. The~number of patches is thus reduced in the relaxation process. Furthermore, the~number of patches decreases as $N$ is increased, as~shown in Figure~\ref{fg:ch7_BCP_chain_length}. This can be understood by noticing that the domain size is larger when the block length is longer. As~$N$ increases, the~critical value of $\epsilon_{AB}$ at which ODT occurs for the diblock copolymers decreases. For~example, at~a melt density of $0.85/\sigma^3$, ODT occurs at $\epsilon_{AB}=1.85\epsilon$ for $N=40$ and $1.28\epsilon$ for $N=100$~\cite{Grest1996}. Since we fix $\epsilon_{AB}$ at $2.0\epsilon$, the~system enters and moves deeper into the ordered region of the phase diagram as the block size increases. This explains the trend that the stripes and patches grow larger and their boundaries sharpen as $N$ increases. For~$N=96$, the~final polymeric nanoparticle after relaxation has a surface pattern that features just a few large stripes and patches. Our results indicate that the chain length is a useful parameter to tune for controlling the surface pattern, including the number of patches, of~a polymeric particle produced via drying a solution droplet of diblock~copolymers.

Experimentally, the~drying of solution droplets discussed above can be realized if a solution is broken into droplets that are suspended and dried in air~\cite{Higuchi2008}. Another technique that can be used to produce polymeric particles or nanoparticle clusters from solution droplets is flash nanoprecipitation (FNP) invented by Johnson and Prud'homme~\cite{Johnson2003a, Johnson2003b}, in which a solute or a mixture of solutes (e.g., drug molecules, nanoparticles, and~polymers) is first dissolved in a solvent and then the solution is rapidly mixed with a non-solvent for the solute(s). In~this process, the~solution is broken into small droplets, which are dispersed in the non-solvent. As~the solvent and non-solvent are miscible, the~solvent is quickly extracted from the droplets, which shrink rapidly, and~the solutes are compressed into particles or clusters by the surrounding non-solvent. This process is quite similar to the drying of a solution droplet. During~drying, the~solvent leaves the droplet via an evaporation process. In~FNP, the~solvent leaves the droplets via diffusion into the non-solvent. In~both cases, the~surface of the droplet recedes, which is either a liquid-vapor interface or an interface between the droplet and the non-solvent, and~the droplet shrinks. Therefore, it is not surprising that polymeric particles with morphologies similar to some of those in Figures~\ref{fg:ch7_homopolymers} and \ref{fg:ch7_copolymers} were produced using FNP~\cite{Grundy2018}. The moving interface method discussed in this paper can thus be applied to droplets undergoing FNP and to address questions such as how the mixing rate of the solvent and non-solvent affects the structure of the resulting~particles.

\section{Summary and~Discussion} \label{sec:ch7_summary}

In this paper, we have reviewed the method of modeling particle suspensions, polymer solutions, and~their mixtures using an implicit solvent with the liquid-vapor interface mimicked by a potential barrier confining all the solutes in the solvent. Their drying process can be studied with the moving interface method, in~which the location of the liquid-vapor interface, i.e.,~the location of the confining potential barrier's minimum, is moved in a prescribed manner. The~evaporation rate can be tuned by varying the speed at which the interface is moved. Various evaporation patterns, including drying films and droplets, can be realized with an appropriate choice of the way in which the interface (i.e, the~equipotential surface of the confining potential) is moved. For~example, the~interface is flat and is translated along its perpendicular direction when a film is dried while it is spherical and is shrunk radially for a drying~droplet.

With the moving interface method, we have studied the drying behavior of a mixed solution of polymer chains and nanoparticles, a~suspension droplet of a bidisperse mixture of nanoparticles, a~solution droplet of a polymer blend, and~a solution droplet of diblock copolymer chains. A~rich set of structures are formed after drying, including stratified films, core-shell clusters (i.e., radially stratified clusters) of nanoparticles, Janus polymeric particles, core-shell polymeric particles, and~patchy polymeric particles. These structures are consistent with those observed previously in explicit solvent simulations and experiments, indicating that the moving interface method with an implicit solvent model can be used in certain situations to yield realistic results. Since the solvent is not treated explicitly, such method has the advantage of significantly reducing the number of particles needed to describe a physical system and thus allows the modeling of much larger systems over longer times and at slow evaporation rates that may be directly comparable with those used in~experiments.

Caution needs to be taken about where the moving interface method can be applied. In~this method, the~solvent is treated as a uniform, viscous, and~isothermal medium. Therefore, the~solvent remains as a background during drying and does not exhibit any flow. The~moving interface method is thus only applicable to situations when the solvent flow is not a crucial factor. For~drying films, spherical droplets, and~cylindrical droplets, this condition can be satisfied and the systems can be modeled with the moving interface method. Examples include a liquid film generated by dip coating~\cite{Javaid2022}, respiratory droplets suspended in air~\cite{Ge2021}, solution droplets created by a spray drying process~\cite{Rodrigues2020}, and a polymer fiber during electrospinning~\cite{Zhou2019}. However, for~systems in which the solvent develops flow patterns during drying, the~moving interface method cannot be directly employed. One example is the drying of a sessile droplet on a substrate, in~which a capillary flow emerges during evaporation, transporting solutes to the edge of the droplet and leading to the famous coffee-ring effect if the peripheral of the droplet is pinned~\cite{Deegan1997,Hu2006}. Since the capillary flow is not captured by an implicit solvent model, it is impossible to produce the coffee-ring deposits with the simplest moving interface method. However, in~these situations the moving interface method can be combined with other techniques such as lattice-Boltzmann method that is able to describe a flow field to study the drying process~\cite{Joshi2010}. Furthermore, there are still many scenarios in which the flow field of the solvent is not a dominant factor and the moving interface method discussed here can be useful because of its computational~efficiency.

In the current implementation of the moving interface method, the~liquid-vapor interface is moved to simulate the drying process but its shape remains unchanged. Therefore, the~method cannot be directly applied to systems where the liquid-vapor interface develops instabilities during evaporation. For~example, when a spin-coated polymeric film is dried, it may develop Rayleigh–B\'{e}nard–Marangoni convective instabilities that cause the evolution of surface morphology of the film~\cite{Chapman2021}. To capture these effects, the~moving interface method needs to be extended to include a model for the liquid-vapor interface that allows the interface to deform and exhibit various instabilities during the drying~process.

Another potential issue with the moving interface method is regarding the dynamics of polymer chains in an implicit solvent. When the Langevin dynamics is applied to each monomer, a~polymer chain essentially follows the Rouse dynamics with the implicit solvent effectively draining through the chain. It is understood that in a dilute polymer solution, the~chain dynamics are better described by the Zimm model, where hydrodynamic interactions make a chain and the solvent in its pervaded volume to behave as a solid object moving through the surrounding solvent~\cite{RubinsteinColby2003}. The previous simulations by Statt~et~al. have revealed that an implicit solvent model of polymer solutions ignoring hydrodynamic interactions could yield outcomes of evaporation even qualitatively inconsistent with those from an explicit solvent model that is matched to the implicit one as far as the equilibrium solution properties are concerned~\cite{Statt2018}. To address this issue, new implicit solvent models need to be developed to yield polymer dynamics matching those in an explicit solvent. Such models can then be used with the moving interface method to study the drying process of polymer-containing soft matter solutions and produce results that agree with those from experiments and explicit-solvent~simulations.

\subsection*{ACKNOWLEDGMENTS}

This material is partially based upon work supported by the National Science Foundation under Grant No. DMR-1944887. The authors acknowledge Advanced Research Computing at Virginia Tech for providing computational resources and technical support that have contributed to the results reported within this paper. S.C. also gratefully acknowledges the support of NVIDIA Corporation with the donation of the Tesla K40 GPUs used for this research. This research used resources of the National Energy Research Scientific Computing Center (NERSC), a U.S. Department of Energy Office of Science User Facility operated under Contract No. DE-AC02-05CH11231. These resources were obtained through the Advanced Scientific Computing Research (ASCR) Leadership Computing Challenge (ALCC). This work was performed, in part, at the Center for Integrated Nanotechnologies, an Office of Science User Facility operated for the U.S. Department of Energy Office of Science. Sandia National Laboratories is a multimission laboratory managed and operated by National Technology and Engineering Solutions of Sandia, LLC., a wholly owned subsidiary of Honeywell International, Inc., for the U.S. Department of Energy's National Nuclear Security Administration under contract DE-NA0003525. This paper describes objective technical results and analysis. Any subjective views or opinions that might be expressed in the paper do not necessarily represent the views of the U.S. Department of Energy or the United States Government.



\end{document}